\documentclass[twoside,11pt]{article}

\usepackage{jmlr2e}
\usepackage{cleveref}
\usepackage{xcolor}
\usepackage{booktabs}
\usepackage{soul}
\usepackage{wrapfig}

\newcommand{\audeep}{\textsc{auDeep}}

\definecolor{pink}{rgb}{0.7,0.0,0.6}

\firstpageno{1}

\begin{document}

\title{auDeep: Unsupervised Learning of Representations 
from Audio with Deep Recurrent Neural Networks}

\author{\name Michael Freitag\textsuperscript{1} \email freitagm@fim.uni-passau.de \\
       \name Shahin Amiriparian\textsuperscript{1,2} \email shahin.amiriparian@tum.de \\
       \name Sergey Pugachevskiy\textsuperscript{1} \email pugachev@fim.uni-passau.de \\
       \name Nicholas Cummins\textsuperscript{1,2} \email nicholas.cummins@ieee.org \\
       \name Bj\"{o}rn Schuller\textsuperscript{1,2,3} \email schuller@ieee.org \\
       \addr \textsuperscript{1}Chair of Complex \& Intelligent Systems\\
       \textsuperscript{\phantom{1}}Universit\"at Passau, 94032 Passau, Germany \\
       \addr \textsuperscript{2}Chair of Embedded Intelligence for Health Care \& Wellbeing\\
       \textsuperscript{\phantom{2}}Augsburg University, Augsburg, Germany \\
       \addr \textsuperscript{3}GLAM -- Group on Language, Audio \& Music \\
       \textsuperscript{\phantom{3}}Imperial College London, London, UK\\}


\maketitle

\begin{abstract}
 \audeep{} is a Python toolkit for deep unsupervised representation learning from acoustic data. It is based on a recurrent sequence to sequence autoencoder approach which can learn representations of time series data by taking into account their temporal dynamics. We provide an extensive command line interface in addition to a Python API for users and developers, both of which are comprehensively documented and publicly available at \url{https://github.com/auDeep/auDeep}. Experimental results indicate that \audeep{} features are competitive with state-of-the art audio classification.
\end{abstract}


\section{Introduction}
Machine learning approaches for audio processing commonly operate on a variety of hand-crafted features computed from raw audio signals. Considerable effort has been put into developing high-performing feature sets for specific tasks.
Recently, representation learning, in particular deep representation learning, has received significant attention as a highly effective alternative to using such conventional feature sets~\citep{Bengio2013,Schmitt2017}. These techniques have been shown to be superior to 
feature engineering for a plethora of tasks, including speech recognition and music transcription~\citep{Bengio2013, Amiriparian2016}. 
Sequential data such as audio, however, poses challenges 
for deep neural networks, as they typically require inputs of fixed dimensionality. In this regard, sequence to sequence learning with \emph{recurrent neural networks} (RNNs) has been proposed in machine translation, for learning fixed-length representations of variable-length sequences~\citep{Sutskever2014}.


In this paper, we present \audeep{}, a first of its kind \textsc{TensorFlow} 
-- based Python toolkit for deep unsupervised representation learning from acoustic data. This is achieved using a recurrent sequence to sequence autoencoder approach. \audeep{} can be used both through its Python API as well as through an extensive command line interface. 


\section{Recurrent Sequence to Sequence Autoencoders}
\label{sec:feature-learning}
Our implementation of sequence to sequence autoencoders extends the RNN encoder-decoder model proposed by \cite{Sutskever2014}. The input sequence is fed to a multilayered \emph{encoder} RNN which collects
key information about the input sequence in its hidden state. The final hidden state of the encoder RNN is then passed through a fully connected layer, the output of which is used to initialise the hidden state of the multilayered \emph{decoder} RNN. The function of the decoder RNN is to reconstruct the input sequence based on the information contained in the initial hidden state.
The network is trained to minimise the root mean squared error between the input sequence and the reconstruction. 
In order to accelerate model convergence, the expected decoder output from the previous step is fed back as the input into the decoder RNN~\citep{Sutskever2014}. Once training is complete, the activations of the fully connected layer are used as the representation of an input sequence.

For the purposes of representation learning from acoustic data, we train sequence to sequence autoencoders built of \emph{long short-term memory} cells or \emph{gated recurrent units} on spectrograms, which are viewed as time dependent sequences of frequency vectors. Two of the key strengths of this approach are (i) fully unsupervised training, and (ii) the ability to account for the temporal dynamics of sequences. 

\section{System Overview}
\label{sec:system-overview}
\audeep{} contains at its core a high-performing implementation of sequence to sequence autoencoders which is not specifically constrained to acoustic data. Based on this domain-independent implementation, we provide extensive additional functionality for representation learning from audio. 

\begin{figure}
\centering
\includegraphics[width=1\textwidth]{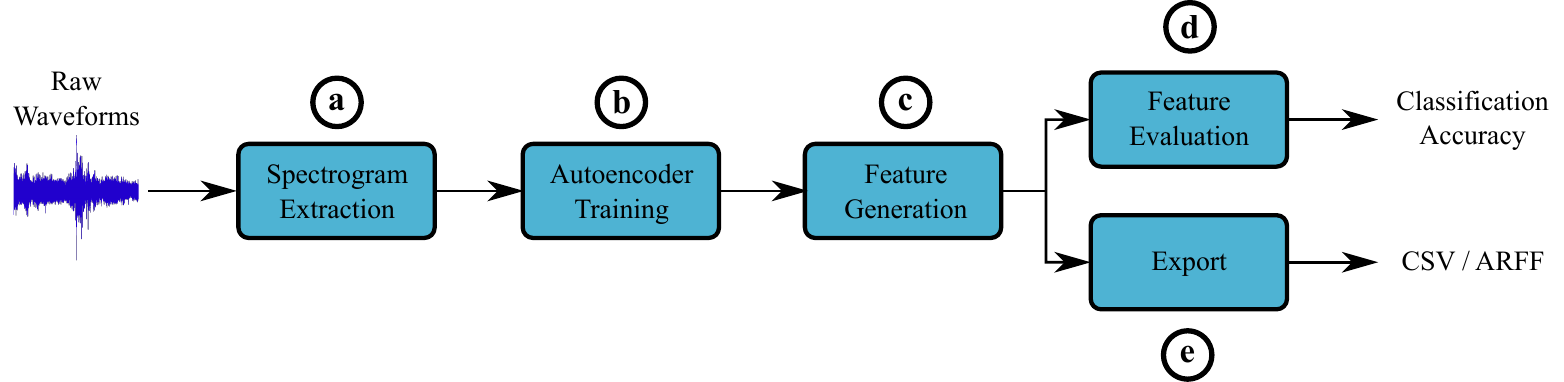}
\caption{Illustration of the feature learning procedure with \audeep{}. A detailed description of the procedure is given in \Cref{sec:practical-usage}.
}
\label{fig:system-overview}
\vspace{-15pt}
\end{figure}
\vspace{-0.2cm}

\subsection{Practical Usage}
\label{sec:practical-usage}
An illustration of 
the feature learning procedure with \audeep{} is shown in \Cref{fig:system-overview}. First, spectrograms are extracted from raw audio files (cf.\,\Cref{fig:system-overview}a). 
Then, a sequence to sequence autoencoder, as previously described, is trained on the extracted spectrograms (cf.\,\Cref{fig:system-overview}b), and the learned representation of each instance is extracted as its feature vector (cf.\,\Cref{fig:system-overview}c). If instance labels are available, a classifier can then be trained and evaluated on the extracted features (cf.\,\Cref{fig:system-overview}d). Finally, the extracted features, and any associated metadata, can be exported to CSV or ARFF for further processing, such as classification with alternate algorithms  (cf.\,\Cref{fig:system-overview}e).

While our system is capable of learning representations of the extracted spectrograms entirely without additional metadata, we provide the possibility to parse instance labels, data set partitions, or a cross-validation setup from a variety of common formats. If available, metadata is stored alongside the extracted spectrograms, and can be used, e.\,g.\ for evaluation of a classifier on the learned representations.
To demonstrate the strength of the learned representations two conventional classifiers are built into the toolkit: a \emph{Multilayer Perceptron} (MLP) with softmax output, and an interface to LibLINEAR~\citep{Fan2008}.

\subsection{Design}
\audeep{} provides a highly modularised Python library 
for deep unsupervised representation learning from audio. The core sequence to sequence autoencoder models are implemented using 
\textsc{TensorFlow}. This implementation substantially extends the built-in sequence to sequence learning capabilities of \textsc{TensorFlow}; for example, the RNNs allow probabilistic feedback and are also reusable, self-contained modules. Furthermore, diversely structured data sets are handled by the system in a unified way without requiring time-consuming manual adjustments.
The topology and parameters of autoencoders are stored as \textsc{TensorFlow} checkpoints which can be reused 
in other applications. 
This enables users, e.\,g.\ to pretrain the encoder RNN with \audeep{} and subsequently apply custom retraining. Data sets are represented in binary format using the platform-independent NetCDF standard, but we provide tools for converting data between NetCDF and CSV/ARFF.

Users are given fine-grained control over the representation learning process through the Python API and the command line interface. 
The system is platform-independent, and has been tested on Windows and various Linux distributions on desktop PCs and in a cluster environment. \audeep{} is capable of running on CPU only, 
and GPU-acceleration is leveraged automatically when available. 



\section{Experiments}
\label{sec:experiments}
\begin{figure}
\centering
\includegraphics[width=0.9\linewidth]{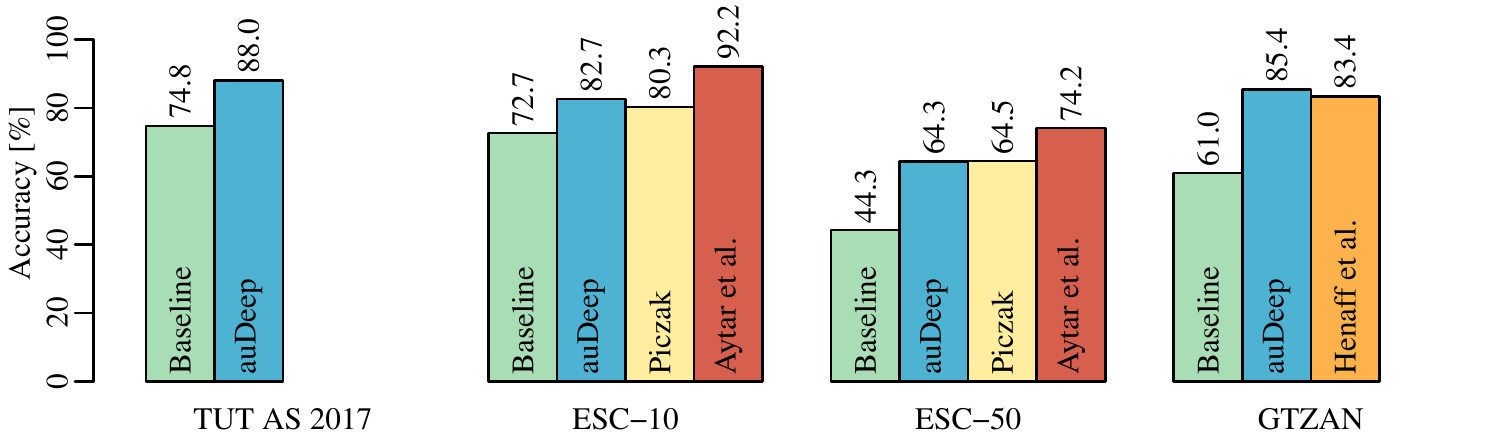}
\caption{Comparison of \audeep{} to baselines and state-of-the-art on four data sets.}
\label{fig:results}
\end{figure}
We demonstrate the 
capabilities of \audeep{} on three audio classification tasks. First, we perform acoustic scene classification on the development partition of the TUT Acoustic Scenes 2017 (TUT AS 2017) data set~\citep{Mesaros2017}. 
Furthermore, we conduct environmental sound classification (ESC) on the ESC-10 and ESC-50 data sets~\citep{Piczak2015ESC}, and, finally, we perform music genre classification on the GTZAN data set~\citep{Tzanetakis2002}. 



We train multiple autoencoder configurations using \audeep{}, and perform feature-level fusion of the learned representations. 
The fused representations are evaluated using 
the built-in MLP with the same cross-validation setup as used for the baseline systems on the TUT AS 2017, ESC-10, and ESC-50 data sets. 
For GTZAN, no predefined cross-validation setup is available, 
therefore we randomly generate five stratified cross-validation folds 
(cf.\,\Cref{fig:results}, \audeep{}). Due to space limitations, we refrain from 
detailing our full experimental setup. 
However, to ensure reproducibility, we distribute the codes and parameter choices used for these experiments 
with \audeep{}.



Finally, we compare the performance of \audeep{} with baseline and state-of-the-art approaches for the different datasets (cf.\,\Cref{fig:results}, identified by authors' names). We observe that \audeep{} either matches or outperforms a \emph{convolutional neural network} approach~\citep{Piczak2015CNN} and a representation learning approach for ESC-10 and ESC-50, and a sparse coding approach for GTZAN~\citep{Henaff2011}. The SoundNet system~\citep{Aytar2016} did achieve stronger performances on ESC-10 and ESC-50. However, this is not a straightforward comparison, as \audeep{} was trained using ESC-10 and ESC-50 data only whilst SoundNet was pre-trained on an external corpus of 2+ million videos. For further details and comparisons with state-of-the-art for the TUT Acoustic Scenes 2017 corpus, the reader is referred to~\cite{Amiriparian17-STS}.






\section{Conclusions}
\label{sec:conclusion}
\audeep{} is an easy-to-use, 
open-source toolkit for deep unsupervised representation learning from audio with competitive performance on various audio classification tasks. Our long-term goal is to grow \audeep{} into a general-purpose deep audio toolkit, by integrating other deep representation learning algorithms such as conditional variational sequence to sequence autoencoders or \emph{deep convolutional generative adversarial networks},
and by extending the feature learning functionality to regression tasks on continuously labelled data.




\acks{\begin{wrapfigure}{l}{0.18\textwidth}
\vspace{-25pt}
  \begin{center}
    \includegraphics[width=0.17\textwidth]{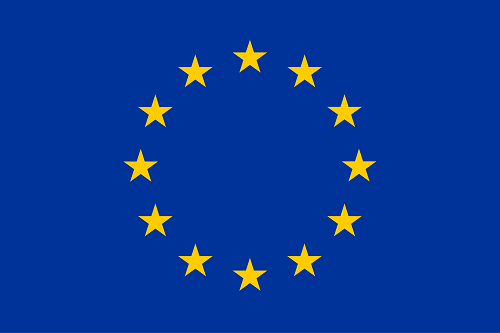}
  \end{center}
\vspace{-25pt}
\end{wrapfigure} The research leading to these results has received funding 
from the Innovative Medicines Initiative 2 Joint Undertaking under grant agreement No\,115902. This Joint Undertaking receives support from the European Union's Horizon 2020 research and innovation programme and EFPIA.}

\vskip 0.2in
\clearpage
\bibliography{refs}

\end{document}